# Estimating terrestrial uranium and thorium by antineutrino flux measurements

**Stephen T. Dye**[1,2] **& Eugene H. Guillian**[3]

[1]Department of Physics and Astronomy, University of Hawaii at Manoa, 2505 Correa Road, Honolulu, Hawaii 96822 USA

[2]College of Natural Sciences, Hawaii Pacific University, 45-045 Kamehameha Highway, Kaneohe, Hawaii 96744 USA

[3]Department of Physics, Engineering Physics, and Astronomy, Queen's University, Kingston, ON, Canada K7L 3N6

**Abstract**

Uranium and thorium within the Earth produce a major portion of terrestrial heat along with a measurable flux of electron antineutrinos. These elements are key components in geophysical and geochemical models. Their quantity and distribution drive the dynamics, define the thermal history, and are a consequence of the differentiation of the Earth. Knowledge of uranium and thorium concentrations in geological reservoirs relies largely on geochemical model calculations. This research report describes the methods and criteria to experimentally determine average concentrations of uranium and thorium in the continental crust and in the mantle using site-specific measurements of the terrestrial antineutrino flux. Optimal, model-independent determinations involve significant exposures of antineutrino detectors remote from nuclear reactors at both a mid-continental and a mid-oceanic site. This would require major, new antineutrino detection projects. The results of such projects could yield a greatly improved understanding of the deep interior of the Earth.

**Introduction**

The production, expected flux, and detection of electron antineutrinos from the decay of long-lived isotopes of uranium, thorium, and potassium in the Earth have been discussed for decades (1-5). Recently, the energies but not the directions of terrestrial antineutrinos have been observed utilizing the inverse β-decay reaction on free protons in a large, monolithic, scintillating liquid detector located on an island arc (6). With this initial observation, non-constraining measurements of the electron antineutrino flux from $^{238}$U and $^{232}$Th decay and of radiogenic heat production within the Earth have been provided. Subsequently, observational neutrino geophysics has been examined at an international workshop (7) and developed in recent reports (8, 9). The present report reviews the research area and proposes a specific neutrino observational program.

Recent reports (8, 9) offer model-dependent predictions of the terrestrial antineutrino flux at potential observation sites, with particular attention to the site of the initial observation (6). Predictions assume the flux originates from uranium and thorium in the crust and mantle only and not the core. Near-surface geological measurements and a whole-Earth geochemical model determine the flux contributions from the crust and mantle, respectively. Comparing the predicted and measured fluxes at a particular site tests the model.

This research develops an alternative method for testing geochemical models. It follows previous work on multi-detector antineutrino spectroscopy (10). It reports experimental criteria for model-independent estimates of the average concentrations of uranium and thorium in Earth's mantle and continental crust. These lead to assessments of radiogenic heating and thorium to uranium ratios in those reservoirs. Such estimates require measurement of the terrestrial antineutrino flux at two geologically distinct locations, ideally a mid-oceanic site and a mid-continental site. Site-dependent fluxes and detector exposures determine uncertainty in the average concentration estimates. Assuming the flux from a reference model (11) allows calculation of benchmark exposures at each site for achieving estimates of a given precision. These exposures set the scale for terrestrial antineutrino detection projects to provide model-constraining estimates of average uranium and thorium concentrations in the mantle and continental crust.

The present method of detecting terrestrial antineutrinos (6) is essentially the same as that used to discover the neutrino (12). Spatially coincident signals from a prompt positron and a delayed neutron efficiently identify electron antineutrino absorption on a free proton. This method, which is subject to the interaction threshold energy of 1.8 MeV, measures well antineutrino energy but not direction (13). The β-decay endpoint energy of $^{40}$K falls below threshold requiring new, unidentified techniques for detection. Four isotopes in the decay series of $^{238}$U and $^{232}$Th have β-decay endpoint energies above threshold: $^{228}$Ac (2.07 MeV) and $^{212}$Bi (2.25 MeV) in the $^{232}$Th series; and $^{234}$Pa (2.27 MeV) and $^{214}$Bi (3.27 MeV) in the $^{238}$U series. Terrestrial antineutrinos in the energy region above 2.3 MeV are due solely to $^{214}$Bi. Given adequate detector energy resolution and calibration this feature permits separate measurement of the flux from $^{238}$U and from $^{232}$Th (10, 14). Due to the weakness of neutrino interactions a convenient unit of detector exposure is $10^{32}$ free-proton years, corresponding to operation of somewhat over 1 kton of scintillating liquid for one year.

Background to the terrestrial antineutrino signal is due to reactor antineutrinos, radioactive contamination of the scintillating liquid, and spallation products from cosmic-ray muons (6). Detectors located at least 1000 km from the nearest nuclear power plant, filled with purified scintillating liquid, and shielded by an overburden equivalent to at least 3000 m of water reduce background to an acceptable level.

The initial observation of terrestrial antineutrinos (6) motivates additional detailed studies and emphasizes the necessity for refined experimental conditions. Further observations from mid-continental and mid-oceanic sites with larger detectors remote from nuclear reactors offer advantages for geological investigations. Identifying these sites allows specific illustrations of geological information resulting from terrestrial antineutrino flux measurements. The selected mid-continental site is in North America (15) and the mid-oceanic site is in the Pacific (16). These sites have an overburden equivalent to more than 4000 m of water and receive a manageable reactor antineutrino flux.

The terrestrial antineutrino flux detected at a given site depends on the quantity and distribution of $^{238}$U and $^{232}$Th within the Earth. Direct sampling of these elements throughout the Earth is not possible with current technology. Estimates of global power derived from surface heat flow measurements (17, 18) provide an upper bound to radioactive heating (19) of a cooling Earth. These limit the quantity of $^{238}$U, $^{232}$Th, and $^{40}$K but do not specify their distribution. Geology provides models for their distribution while allowing considerable variability (11). Geochemists calculate $^{238}$U and $^{232}$Th concentrations in the relatively thin continental crust roughly two orders of magnitude greater than in the thick mantle and values at or near zero in the core (20, 21). This generates a terrestrial antineutrino flux about four times greater on continents than in oceans. A comparison of the measured flux at these sites tests this fundamental geochemical prediction. If valid, oceanic detectors need to be several times larger than continental detectors to achieve comparable event rates.

**Results**

Consider measuring at a single site the terrestrial antineutrino flux given by a reference model and its uncertainties (11). According to the reference model the mid-continental site receives 82% of the flux from continental crust with the majority coming from within about 600 km of the detector. In contrast, the mid-oceanic site receives 72% of the flux from the mantle with the majority coming from within about 5000 km of the detector (22). Without directional information, estimates of flux contributions from continental crust and mantle rely on model-specific information and uncertainties. Determining the contribution from one reservoir requires subtracting from the total flux the *model-predicted* contribution from the other reservoir.

Table 1 presents percentage errors in predicted terrestrial antineutrino fluxes arising from systematic uncertainties in the reference model (11). The mid-continental site affords useful, model-dependent estimates of uranium and thorium in the continental crust but not in the mantle. Similarly, the mid-oceanic site offers precise, model-dependent estimates of uranium and thorium in mantle but not in continental crust. This precision results from the relatively small model uncertainties in uranium and thorium concentrations in continental crust.

A flux measurement at a single site estimates globally averaged $^{238}$U and $^{232}$Th concentrations. It does not provide a unique solution to the distribution of uranium and thorium within the Earth nor to the radiogenic heating. An analysis combining flux measurements from two geologically distinct sites, one mid-continental and the other mid-oceanic, most effectively isolates flux contributions from continental crust and mantle. Now the contribution from one reservoir results from subtracting from the total flux the *measured* contribution from the other reservoir. This removes systematic uncertainties inherent in model-dependent single site estimates. This analysis develops multi-detector terrestrial antineutrino spectroscopy (10).

Assumptions remaining in this analysis include negligible antineutrino flux from the core and a potentially rich layer of uranium and thorium surrounding the core, and the sampled portion of a reservoir is representative of the entire reservoir. Recent laboratory experiments place an upper bound on the concentration of uranium in the core (23, 24) supporting a negligible flux from this reservoir. Testing for lateral bulk heterogeneities in the two reservoirs requires multiple measurement sites. For continental crust this entails more than one observatory. For mantle, however, a mobile oceanic observatory facilitates multiple deployments (25).

The precision of isotope concentration estimates resulting from multi-detector terrestrial antineutrino spectroscopy scales with the square root of detector exposures. Figure 1 presents contours of 20% uncertainty on estimates of the concentrations of $^{238}$U and $^{232}$Th in mantle and continental crust. Concentration estimates enable calculation of the thorium to uranium ratio and the radiogenic heating in a two reservoir Earth system. Due to the terrestrial antineutrino spectrum the analysis determines $^{238}$U concentrations better than $^{232}$Th concentrations. The axes of Figure 1 are exposures at the continental and oceanic sites. A benchmark is the intersection of the radiogenic heating contours for the mantle and continental crust. This intersection corresponds to exposures of $8 \times 10^{32}$ free-proton years at the mid-continental site and $5 \times 10^{33}$ free-proton years at the mid-oceanic site. This pair of exposures offers estimates of the total radiogenic heat and average $^{238}$U concentrations in the two reservoirs to better than 20%. Achieving these benchmark exposures within a five-year timeframe suggests a continental detector in the 2-4 kton range and an oceanic detector in the 15-30 kton range.

4Figure 2 illustrates how combining terrestrial antineutrino flux measurements from the mid-continental and mid-oceanic sites determines $^{238}$U and $^{232}$Th concentrations in continental crust and mantle. A flux measurement from one site defines a broad range of possible concentrations in the two reservoirs. A higher concentration of uranium or thorium in one reservoir requires a lower concentration in the other. This linear relationship depends on the site. Flux measurements from two sites, especially geologically distinct sites providing sufficiently different relationships, specify concentration estimates by the intersection of the ranges.

Different geological models predict different concentrations of uranium, thorium, and potassium in the various Earth reservoirs (28, 29). Models with higher concentrations produce correspondingly higher antineutrino fluxes and radiogenic heating. Flux measurements resulting from the benchmark exposures compare the reference model (11) with geological models with higher isotope concentrations (28). The reference model supplies 11 TW of heating to the mantle compared with 18 TW and 26 TW for the other models. Table 2 presents the significance of the resolution when comparing measurements of uranium and thorium concentrations in the mantle and continental crust. The resolution is greatest for the measurement of uranium concentration in the mantle.

**Discussion**

Measuring the terrestrial antineutrino flux both from continent and ocean is within current technological capability. Detectors with up to $4 \times 10^{33}$ free proton targets are currently under consideration (30). At this size observations for a combined analysis leading to 10% measurements of the uranium content and <20% measurements of radiogenic heat production of both mantle and continental crust are possible after deployments of just a few years. These direct measurements of the Earth's deep interior would help constrain models of Earth composition and dynamics leading to significant advances in geochemistry and geophysics.

Near term prospects for measuring terrestrial antineutrino flux reside with continuing observations on an island arc (8) and a new observatory on the Canadian shield (31).* Both detectors are ~1 kton in size ($< 4 \times 10^{31}$ free-proton targets) and subject to considerable reactor antineutrino flux, the former more severe than the latter. The Canadian detector should be able to confirm observation of terrestrial antineutrinos after operating for about three years.

Terrestrial antineutrino flux measurements are the only identified, feasible method to experimentally determine the distribution of uranium and thorium in the interior of the Earth. Measurements from two geologically distinct detection sites remote from nuclear reactors provide model-independent estimates of uranium and thorium concentrations in continental crust and mantle. This information is vital for understanding the Earth's geophysical structure and dynamics. If, as expected, these elements are much less concentrated in the mantle than in the continental crust, the oceanic detector needs to be several times larger than the continental detector.

* A small detector in Italy is now taking data but expects little precision in a terrestrial antineutrino flux measurement (32).

**Methods**

Extraction of average uranium and thorium concentrations in Earth reservoirs begins with the terrestrial antineutrino flux: $\Phi_X(\vec{r}_0) = \left(\dfrac{A_X N_X}{2 R_\oplus}\right) \dfrac{R_\oplus}{2\pi} \int dV \dfrac{a_X(\vec{r}) \rho(\vec{r})}{|\vec{r} - \vec{r}_0|^2}$

*X* denotes $^{238}$U, $^{232}$Th. $\vec{r}_0$ ($\vec{r}$) indicates flux measurement site (Earth volume element *dV*). $A_X$ is the isotope activity per gram; $A_U = 1.24 \times 10^4$ Bq/g, $A_{Th} = 4.04 \times 10^3$ Bq/g. $N_X$ is the number of emitted neutrinos per decay; $N_U = 6$, $N_{Th} = 4$. $a_X(\vec{r})$ and $\rho_X(\vec{r})$ specify isotope concentration and density of volume elements, respectively. $R_\oplus$ is the average Earth radius.

Assuming $^{238}$U, $^{232}$Th reside only in continental crust and mantle with uniform concentration, the flux equation becomes:
$\Phi_X(\vec{r}_0) = A_X N_X /(2R_\oplus)[a_{X,CC} I_{CC}(\vec{r}_0) + a_{X,M} I_M]$. $a_{X,CC}$ ($a_{X,M}$) is the average isotope concentration in continental crust (mantle). $I_{CC}(\vec{r}_0)$, $I_M$ are defined as:
$I_S = \frac{R_\oplus}{2\pi} \int_S dV \frac{\rho(\vec{r})}{|\vec{r}-\vec{r}_0|^2}$, where the integral is over reservoir S=CC,M. $I_M = 1.48 \times 10^{18}$ g/cm for surface flux calculations assuming a spherically symmetric mantle. The terrestrial antineutrino flux becomes a linear combination of average isotope concentrations in the reservoirs according to: $\begin{pmatrix} \Phi_X(\vec{r}_1) \\ \Phi_X(\vec{r}_2) \end{pmatrix} = C_X \begin{pmatrix} I_{CC}(\vec{r}_1) & I_M \\ I_{CC}(\vec{r}_2) & I_M \end{pmatrix} \begin{pmatrix} a_{X,CC} \\ a_{X,M} \end{pmatrix}$,
with the quantity $C_X = A_X N_X /(2R_\oplus)$.

Solving these equations requires matrix inversion. Concentration uncertainties depend inversely on the magnitude of the matrix determinant $I_M |I_{CC}(\vec{r}_1) - I_{CC}(\vec{r}_2)|$. Maximizing the difference in crust integrals minimizes concentration uncertainties. This validates selection of a mid-continental ($I_{CC} = 4.07 \times 10^{16}$ g/cm) and mid-oceanic ($I_{CC} = 0.306 \times 10^{16}$ g/cm) pair of sites.

Dividing the spectrum into lower and higher energy regions isolates flux contributions from uranium and thorium. Both uranium and thorium contribute to the lower energy region (1.8-2.3 MeV). Only uranium contributes to the higher energy region (2.3-3.3 MeV). The detected neutrino events in these regions ($N_1 = N_{Th} + (1-f)N_U$ in lower, $N_2 = fN_U$ in higher; $f = 0.541$) determine each region's neutrino flux ($\varphi_1, \varphi_2$). For example, $^{238}$U concentrations in continental crust and mantle obtain from $a_{U,CC} = \dfrac{\varphi_2(\vec{r}_c) - \varphi_2(\vec{r}_o)}{fC_U[I_{CC}(\vec{r}_c) - I_{CC}(\vec{r}_o)]}$ and $a_{U,M} = \dfrac{I_{CC}(\vec{r}_c) \cdot \varphi_2(\vec{r}_c) - I_{CC}(\vec{r}_o) \cdot \varphi_2(\vec{r}_o)}{fC_U I_M [I_{CC}(\vec{r}_c) - I_{CC}(\vec{r}_o)]}$. The vector $\vec{r}_c$ ($\vec{r}_o$) indicates continental (oceanic) site. Similar equations specify $^{232}$Th concentrations in the reservoirs.

**Acknowledgments**

We thank C. Lan for allowing us to use his code to calculate the integral over geological reservoirs and J.G. Learned and W.F. McDonough for discussions and comments.

Figure 1 | Contours show the exposures required at the continental and oceanic sites to achieve 20% uncertainty in estimates of uranium and thorium concentrations and radiogenic heating in a two reservoir Earth system. Calculations include background from all commercial nuclear reactors (26) operating at full power 80% of the time, an antineutrino detection efficiency of 0.7 (6), and a reduction of antineutrino flux by a factor of 0.57 due to best-fit neutrino oscillation parameters (27). Heating estimates assume a ratio of potassium to uranium of $1.2 \times 10^4$ (20).

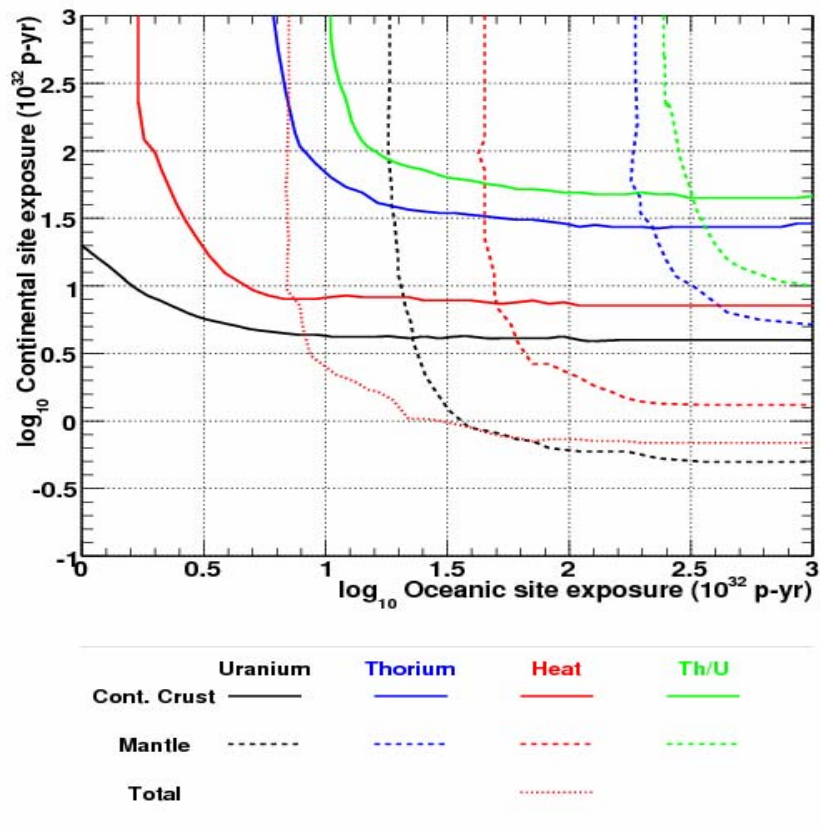

Figure 2 | The range of concentrations of uranium (top panel) and thorium (bottom panel) in the continental crust (horizontal axes) and mantle (vertical axes) from flux measurements at the mid-continental (black lines) and mid-oceanic (blue lines) sites intersect at the specified value. Thin lines represent one sigma errors determined by the benchmark exposure at each site.

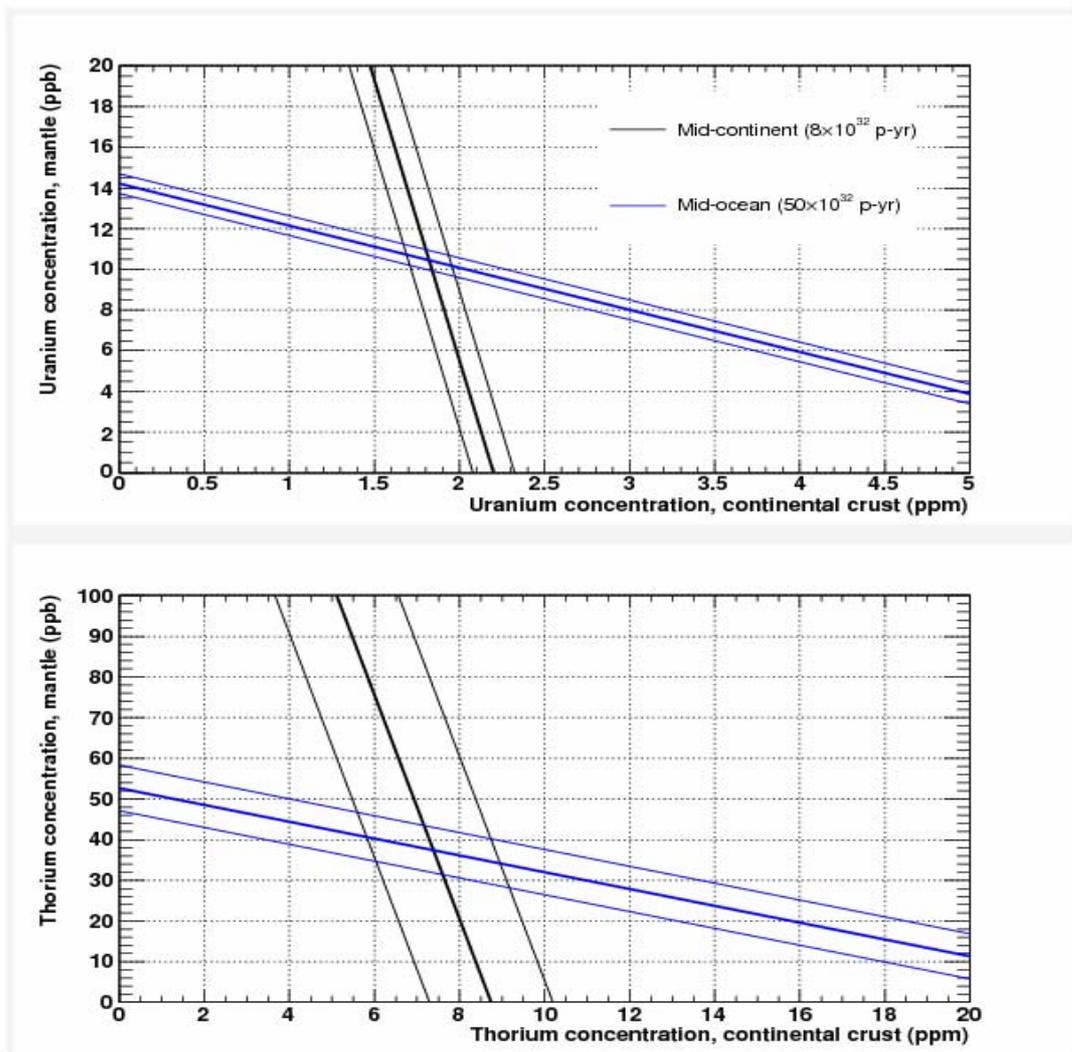

**Table 1 Percentage errors in antineutrino flux resulting from model uncertainties**

Percentage errors on estimates of antineutrino flux from uranium and thorium in mantle and continental crust made at either a mid-continental site or a mid-oceanic site result from reference model uncertainties[11].

|  | Mid-continent | | Mid-ocean | |
|---|---|---|---|---|
|  | Uranium | thorium | uranium | thorium |
| Continental crust | +12/-37 | +14/-36 | +182/-100 | +236/-100 |
| Mantle | +189/-77 | +204/-88 | +12/-5 | +12/-6 |

**Table 2 Resolving geological models**

Significance is given in number of sigma for resolving geological models with different levels of radiogenic heating in the mantle from the reference model[11] using the benchmark exposures.

|  | Continental crust | | Mantle | |
|---|---|---|---|---|
|  | uranium | thorium | uranium | thorium |
| 18 TW | -2.6 | -0.4 | 6.0 | 1.6 |
| 26 TW | -3.3 | -0.2 | 12.2 | 3.4 |